\documentclass[runningheads]{svmult}

\usepackage{makeidx}   
\usepackage{graphicx}  
\usepackage{subeqnar}  
\usepackage{multicol}  
\usepackage{physprbb}  
\makeindex             

\begin{document}
\title*{Mergers and binary systems of SMBH in the contexts of nuclear
activity and galaxy evolution}
\toctitle{Mergers and binary systems of SMBH in the contexts of nuclear
activity and galaxy evolution}
%
%
\titlerunning{Mergers and binary systems of SMBH}
%
\author{Andrei Lobanov}

\authorrunning{Andrei Lobanov}
%
%
\institute{Max-Planck-Institut f\"ur Radioastronomie,
Auf dem, H\"ugel 69, 53121 Bonn, Germany}

\maketitle              

\begin{abstract}
  The dynamic evolution of binary systems of supermassive black holes
  (SMBH) may be a key factor affecting a large fraction of the
  observed properties of active galactic nuclei (AGN) and galaxy evolution. Different classes
  of AGN can be related in general to four evolutionary stages in a
  binary SMBH: 1) early merger stage; 2) wide pair stage; 3) close
  pair stage; and 4) pre-coalescence stage.


\end{abstract}

\noindent
Mergers are expected to occur frequently over the course of galaxy evolution.
Formation of binary (or multiple) systems of SMBH is a likely outcome of
galactic mergers. Binary black hole (BBH) systems should therefore play an
important role in the nuclear activity in galaxies, since the latter is
believed to be closely related to SMBH. 
Analytical work may help identifying main trends of the co-evolution
of AGN and BBH. A good starting point for this work is provided by a
number of studies of BBH dynamics (e.g., [1] and subsequent works) and
interaction of supermassive black holes with nuclear environment in
galaxies [2,4].  Based on these studies, a scheme can be
proposed that
connects distinct stages of the binary evolution to characteristic
types of AGN and galactic morphology associated with them.

\textbf{1.~Early merger --- Low-power AGN:}~Individual galaxies or early mergers (while both BH retain their accretion
disks). Timescale and magnitude of nuclear activity depend on the conditions
in the central regions, and it is likely that cavity is formed around the BH,
producing a ``starving'', low--power AGN [2].  If galaxies were formed at
redshifts $z\sim5-10$, the peak of single BH activity in galaxies is likely to
occur at $z\sim1$.  Around this redshift, AGN with single SMBH may represent a
(probably small) fraction of quasars and FR\,II type radio galaxies.  At later
epochs a single SMBH in the center of a galaxy is expected to reduce its
fueling rate to $F \le 10^{-3}\,[\dot{\mathrm{M}}_\mathrm{Edd}]$, and support
a typical luminosity of $L \le L_\mathrm{ Edd}\, F \approx 10^{43} M_8$\,erg/s.
($M_8 = M_\mathrm{bh}/10^8\,M_{\mathord\odot}$).
This activity would be similar to that found in a typical Seyfert galaxy.
The activity should remain weak during an early merger and relaxation of the
galactic cores, which is expected to last for $\sim 10^8$\,years,
[4,5].  An AGN at this stage would have weak pc-scale and (FR\,I)
kpc-scale jets, weak broad line regions, and very
weak variability.

\textbf{2.~Wide pair --- High-power AGN:}~After the merger, the BHs
sink toward the center of the stellar distribution and form a
gravitationally bound system, with typical orbital separations
$r_\mathrm{ b}\sim 1-10$\,pc and initial orbital velocities
$v_\mathrm{ init} \sim 10-100$\,km/s. The dynamical friction would
reduce the orbital separation to $r_\mathrm{ b} \sim 0.1-1$\,pc, and
the smaller BH would eventually lose its accretion disk.  The
accretion disk is aligned with the orbital plane and disrupted by the
secondary BH. Interaction of the BBH with the stars and gas would
increase the fueling rate by a factor of 10--100 [2], bringing it
close to $\dot{M}_\mathrm{ Edd}$ and supporting a luminosity of
$\le10^{46}$\,erg/s on timescales of $\sim 10^8$\,years. BBH systems
at this stage should produce strong pc-scale and (FR\,II) kpc-scale jets;
strong broad line regions, and  variability on timescales
$\tau_\mathrm{ var}\sim 10^2$--$10^4$\,days.

\textbf{3.~Close pair --- Radio-quiet AGN:}~At orbital separations
$r_\mathrm{ b} \sim 10^{-2}-1$\,pc, the interaction of the secondary
BH with the accretion disk intensifies so that it may even lead to a
complete destruction of the disk.  A turbulent activity in the nuclear
region would result in strong thermal emission in the optical and
high-energy band, varying on timescales $\tau_\mathrm{ var}\sim
10^0$--$10^3$\,days.  The jet production stops, and the level of radio
emission is reduced substantially.  The fueling rate is also reduced,
and the resulting luminosities would reach up to $\le 10^{45}$\,erg/s.
This stage lasts for $\le 10^8$\,years.  An AGN at this stage is
 a ``radio quiet'' QSO (one should remember, however, that
there are several possible factors potentially capable of quenching
the jet production in AGN).

\textbf{4.~Pre-coalescence --- Intraday variable AGN:}~At separations
$r_\mathrm{ b} \le 10^{-2}$\,pc, gravitational radiation becomes the
most important evolutionary factor. With $r_\mathrm{ b}\gg r_\mathrm{
  acc}$, it is possible that an accretion disk is formed again around
the BBH.  Luminosities of up to $\le 10^{45}$\,erg/s should be
expected.  This stage would last for $\sim
10^7$\,years. A typical AGN at this stage would be an intraday
variable source (with prominent variability on timescales
$\tau_\mathrm{ var}\sim 10^{-1}$--$10^3$\,days), with ``re-started''
pc-scale radio jets, (and possible kpc-scale relics), and a prominent
broad line region.  The relativistic effects and orbital motion will
result in
variability on timescales that correspond to brightness temperatures
of up to
\[ T_\mathrm{ b,max} = 5.58\cdot 10^{11} \Delta S_\mathrm{ var} \delta^{2-\alpha}
\left[\frac{\lambda_\mathrm{ obs} D_\mathrm{ L} (\mu^3 + \mu^2)^{1/5}}
{M_{8} (1+z)^2}\right]^2\, [\mathrm{K}]\,,
\]
where $\mu$ is the mass ratio of the binary and $\delta$ is the Doppler factor
of the emitting material. Example: 0917+624: 
$z=1.446$, $\Delta S_\mathrm{ var} = 0.14$\,Jy, $\tau_\mathrm{ min} = 0.285\,d$
at $\lambda_\mathrm{
  obs}=6$\,cm $T_\mathrm{ b,obs} = 1.4\cdot 10^{19}$\,K [3].
Model estimate: $T_\mathrm{ b,max} = 4\cdot
10^{20}$\,K, for $M_\mathrm{ bh}=10^7$M$_{\mathord\odot}$,
$\mu=1$, $\delta=1$.


The proposed connection between BBH evolution and AGN is not necessarily
unique and exclusive, but it should provide a viable skeleton for building up
more complex and detailed models relating nuclear activity to the properties
of multiple black holes in active galaxies. In particular, it would be
important to investigate joint cosmological evolution of active galaxies and
supermassive BBH embedded into their nuclei.

%


\begin{thebibliography}{8.}
\addcontentsline{toc}{section}{References}

\bibitem{bbr84} M.C. Begelman, R.C. Blandford, M.J. Rees: Nature \textbf{287},
  307 (1980)

\bibitem{dok91} V.I. Dokuchaev: MNRAS, \textbf{251}, 564 (1991)

\bibitem{kra99} A. Kraus et al: A\&A, \textbf{352}, 107 (1999)

\bibitem{pol94} A.G. Polnarev, M.J. Rees: A\&A, \textbf{283}, 301 (1994)

\bibitem{roo91} N. Roos: A\&A, \textbf{104}, 218 (1991)

\end{thebibliography}
\end{document}